# Memory Forensic Analysis of MQTT Devices

Anal Shah
*Institute of Forensic Science*
*Gujarat Forensic Sciences University*
Gandhinagar, India
analshah1705@gmail.com

Palak Rajdev
*Institute of Forensic Science*
*Gujarat Forensic Sciences University*
Gandhinagar, India
palakhrajdev@gmail.com

Jaidip Kotak
*Institute of Forensic Science*
*Gujarat Forensic Sciences University*
Gandhinagar, India
jaidipkotak@gmail.com

***Abstract*** **- Internet of Things is revolutionizing the current era with its vast usage in number of fields such as medicine, automation, home security, smart cities, etc. As these IoT devices' uses are increasing, the threat to its security and to its application protocols are also increasing. Traffic passing over these protocol if intercepted, could reveal sensitive information and result in taking control of the entire IoT network. Scope of this paper is limited to MQTT protocol. MQTT (MQ Telemetry Transport) is a light weight protocol used for communication between IoT devices. There are multiple brokers as well as clients available for publishing and subscribing to services. For security purpose, it is essential to secure the traffic, broker and end client application. This paper demonstrates extraction of sensitive data from the devices which are running broker and client application.**

## I. Introduction

Kevin Ashton first proposed the term Internet of Things explaining it as a system of interconnected devices in 1999. Since then, it has revolutionized the current computing networks by adding physical objects surrounding us on a network with each device being uniquely identifiable and connected to each other [1].

Because of the increasing number of IoT devices in personal environments such as computers and smartphones, opportunities and risks are becoming tangible from a forensic perspective. Forensic investigation can greatly benefit from the traces left by these devices. But with every good side, comes a bad side; these traces can also be used by criminals to undermine sensitive information from device [2].

There are many application level protocols used in IoT devices like Hyper Text Transfer Protocol(HTTP), Constrained Application Protocol (CoAP), Extensible Messaging and Presence Protocol(XMPP), Advanced Message Queuing Protocol(AMQP) and Message Queuing Telemetry Protocol (MQTT). In this paper, we will be discussing the most popular protocol, MQTT. In this protocol, data is published to a particular topic at regular intervals by the sensors (publishers). The devices (subscribers) which have registered to that particular topic will receive the messages from the broker each time that topic receives new message [3].

Along with the network traces and vulnerabilities exposed by the protocol itself, an important location to extract data of any MQTT application is in the application's memory. There will always be traces that are left in the memory and these traces can provide powerful insights into the data that is transferred between the broker and the client devices along with other sensitive information. For criminals, these data can provide powerful ways to further damage the network.

In this paper, we have demonstrated mechanism to used for extracting traces in forensically sound manner from the applications running MQTT protocol, type of information which can be gathered from these traces and how this information can prove helpful to either forensic investigator or an attacker. We have experimented with various brokers as well as client applications on ubuntu virtual machines (versions 16 and 18) and found out the traces left by each one in the memory. The reason for choosing linux operating system is that most of the IoT application is linux based.

## II. Related Work

### A. *MQTT Protocol*

MQTT is a publish/subscribe protocol used in major IoT devices that are constrained in terms of bandwidth, data storage, etc. It is used in areas where communication links supply low throughput like home automation, smart cities, monitoring, SCADA, etc [4]. MQTT is on top of TCP, ultimately benefitting from error and flow control for a single packet to the lower layers of the protocol pyramid. As MQTT works on publisher/subscriber paradigm, information is classified through "topic" hierarchies with path and nodes followed by subscribers [5]. Security is an essential aspect when working with the protocol with the focus on authentication, access control, data integrity and confidentiality. But along with that, we have to consider the devices that are hosting the broker or the client applications. If the device itself is vulnerable, an attacker may be able to gather forensic traces of an application with the help of just taking a memory dump.

### B. *IoT Forensics*

The Internet of Things is growing at a fast pace and is thus creating opportunities for cyber-attacks. IoT devices have sensors or actuators that generate data, which can make them excellent digital witnesses, as they can capture traces of activities of potential use in investigations. Francesco Servida and Eoghan Casey have proposed a work to increase familiarity with traces from IoT devices in a smarthome which can be useful for forensic purposes. This work presents a study of IoT devices and their smarthome applications, providing methods to extracting and analysing digital traces. Most smart homes, and even smart buildings lack any forensic preparedness so this work concentrates on traces that can be obtained from IoT devices at a crime scene and associated smart phones. And also proposing a generalized practical process that extends existing methods for examining smart phones [2].

## III. Need For IoT Forensics

IoT is slowly becoming a part of our everyday life from taking care of the home temperature to smart cars and smart management of the cities, which has ultimately led to increase in the number of cyber security incidents on the IoT devices/applications [6]. Therefore, it has necessitated the collection and analysis of digital evidence from different types of IoT devices i.e. IoT forensics. As the connected devices in the IoT infrastructure can be versatile like sensor devices, communication devices, cloud storage, etc., the approach for forensic analysis of IoT also differs from that of traditional analysis.

The traces of digital evidence can be collected from each and every device from its memory which can ultimately help the attacker to discern the communication between the devices without intercepting the network. For example, if attackers are able to get inside the computers hosting the applications of MQTT, they can get personal information of the people living in home just by examining the memory dump and thus invade privacy.

IoT devices can serve as a gateway to home or corporate network if not secured properly. 80% of the world's data is in private servers and if the attackers are able to gain entry to those servers, the damage can be monumental. While we cannot foresee each and every attack, at least proactive measures should be taken to mitigate them. This includes securing the systems running IoT applications and network security. Security aspects such as privacy and confidentiality of data should be included in the MQTT protocol as IoT security will become increasingly important in preventing business and personal catastrophes [7].

Along with focusing on the IoT sensors and firmware security, care should also be taken of the computers/mobile phones hosting these applications. Due to lack of security measure if an attacker is able to access the devices hosting IoT applications, it is easy to extract the data from the application processes from its memory and thus collect the volatile data which can be equally destructive.

## IV. Methodology

In MQTT protocol, publisher publishing messages and communicating with users or clients subscribing to that topics is ordinarily considered as a publish/subscribe model or methodology. As shown in figure 1, the devices on one end publishes the data to topics on broker and at the opposite end devices getting their messages from subscribed topics.

For our analysis, we have used different brokers and client applications that use MQTT protocol. We ran different broker applications with the same client and to its opposite we ran different client applications against the same broker for noting the behaviour in the same environment. We have used linux virtual machine for testing the results and have analysed the memory dump taken using procdump with the use of hex editor.

## V. Analysis

Based on the setup explained in the methodology, we have divided the analysis into sections: Analysis of brokers and Analysis of clients. Below are the details:

### A. *Analysis of Broker Applications*

For this analysis, 5 different brokers were used against the same client application named mosquitto. Broker was running in a linux virtual machine, mosquitto_sub is used from a different linux machine to subscribe to a topic and mosquitto_pub is used to publish to that topic using the broker. After completion of data transfer memory dump of the broker was taken and was manually examined to understand what data is stored in it. Table 1 describes name of the broker applications and data that was extracted from its memory dump. For Mosca and hbmqtt brokers, their respective clients; mqtt client and hbmqtt clients were used.

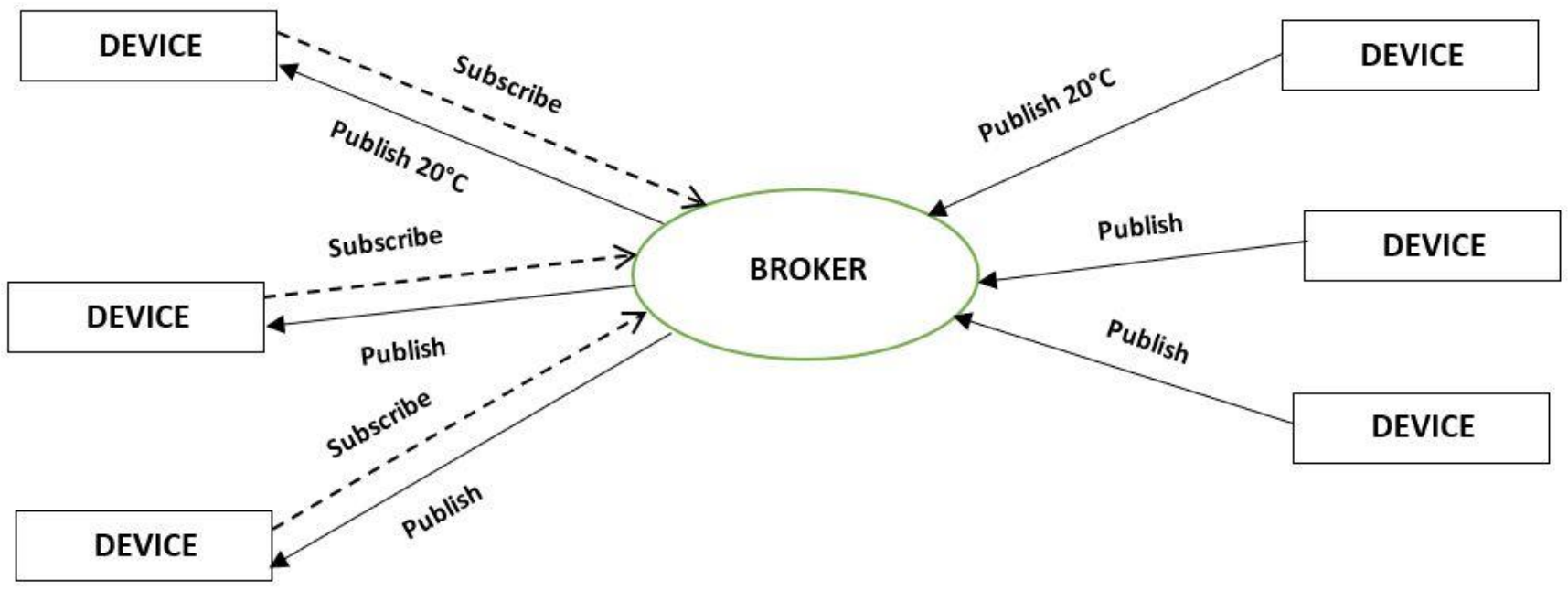

Fig. 1 Basic MQTT Publish/Subscribe Model

TABLE I

FEATURES AND MEMORY DUMP ANALYSIS OF BROKER APPLICATIONS

| Sr. No. | Type of data extracted from process dump | Eclipse Mosquitto | Mosca | Hbmqtt | Rabbit MQ | Hive MQ |
|---|---|---|---|---|---|---|
| 1 | Client id (eg: mosqsub/12-ubuntu) | ✓ | ✓ | | ✓ | ✓ |
| 2 | IP address of broker | | ✓ | ✓ | ✓ | |
| 3 | IP address of client | ✓ | | ✓ | ✓ | ✓ |
| 4 | Topic name | ✓ | ✓ | ✓ | ✓ | ✓ |
| 5 | Message sent by subscriber | ✓ | ✓ | ✓ | ✓ | ✓ |
| 6 | Location from where the broker application is running in the VM | ✓ | ✓ | ✓ | ✓ | ✓ |

TABLE II

FEATURES AND MEMORY DUMP ANALYSIS OF CLIENT APPLICATIONS

| Sr. No. | Type of data extracted from process dump | Mosquitto | Paho MQTT client | MQTT fx |
|---|---|---|---|---|
| 1 | Message sent by subscriber | ✓ | | ✓ |
| 2 | Ip address of broker | ✓ | | |
| 3 | Location from where the broker application is running in the VM | ✓ | ✓ | ✓ |
| 4 | Topic name | ✓ | | ✓ |
| 5 | Logged in user details (username , password) | | ✓ | |

### *B. Analysis of Broker Communicating over TLS/SSL*

As a part of research we also analysed the brokers like Mosquitto, Mosca and Hive MQ running on port 8883 and which uses TLS encryption. Although the traffic travelling between the broker and client was secure due to encryption, if one could gain access to the system running the broker application, one could still gain a lot of knowledge about the communication between the broker and the clients.

Topic name, message sent by the subscriber, client id, location from where the broker application is running in the virtual machine, etc. was all recovered from the memory dump of the broker application.

Even though network capturing tools could not display the contents of the packets due to TLS layer, one could still read the contents from the memory dump of the broker. Thus, it proves that along with securing the traffic, one should also secure systems which is running the broker application.

** Results are same which is displayed in Table-1 for brokers communicating over SSL.*

### *C. Analysis of Client Applications*

For analysis of clients of MQTT, three linux based open source clients were used. Mosquitto broker was running on one VM machine and on the other end clients were running for publishing and subscribing. Three different clients memory dump was taken from subscriber side by procdump tool and was examined using hex editor to see to understand what data resides in memory while client gets connected. Table 2 describes name of the client applications data that was extracted from its memory dump.

## VI. CONCLUSION AND FUTURE WORK

IoT protocol examined in this paper was MQTT protocol. We performed memory forensic analysis of different broker and client applications, examined the artifacts discovered in the dump and also discussed the need for IoT forensics as well as the need for securing the systems which are running broker or client applications.

As a part of our future work, we will continue to discover artifacts using forensic analysis for other IoT protocols.

## APPENDIX: SNIPPETS OF DATA FROM MEMORY

1. Below is the data from memory containing the whole mqtt subscribe-publish packet details when a process dump of Mosquitto broker was analysed.

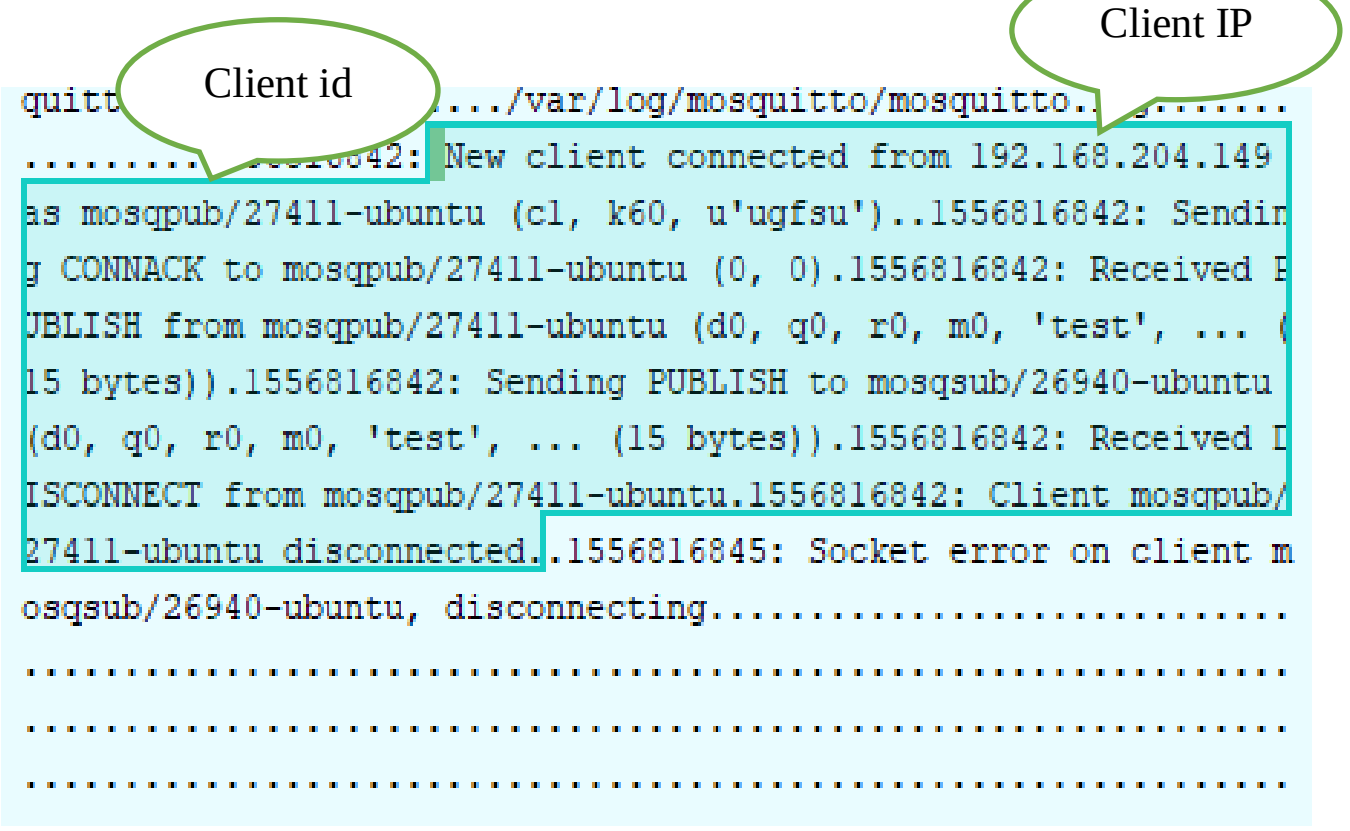

Fig. 2 Mosquitto Broker memory dump example

2. Various configuration information can be found out in the memory dump of the Rabbit MQ broker.

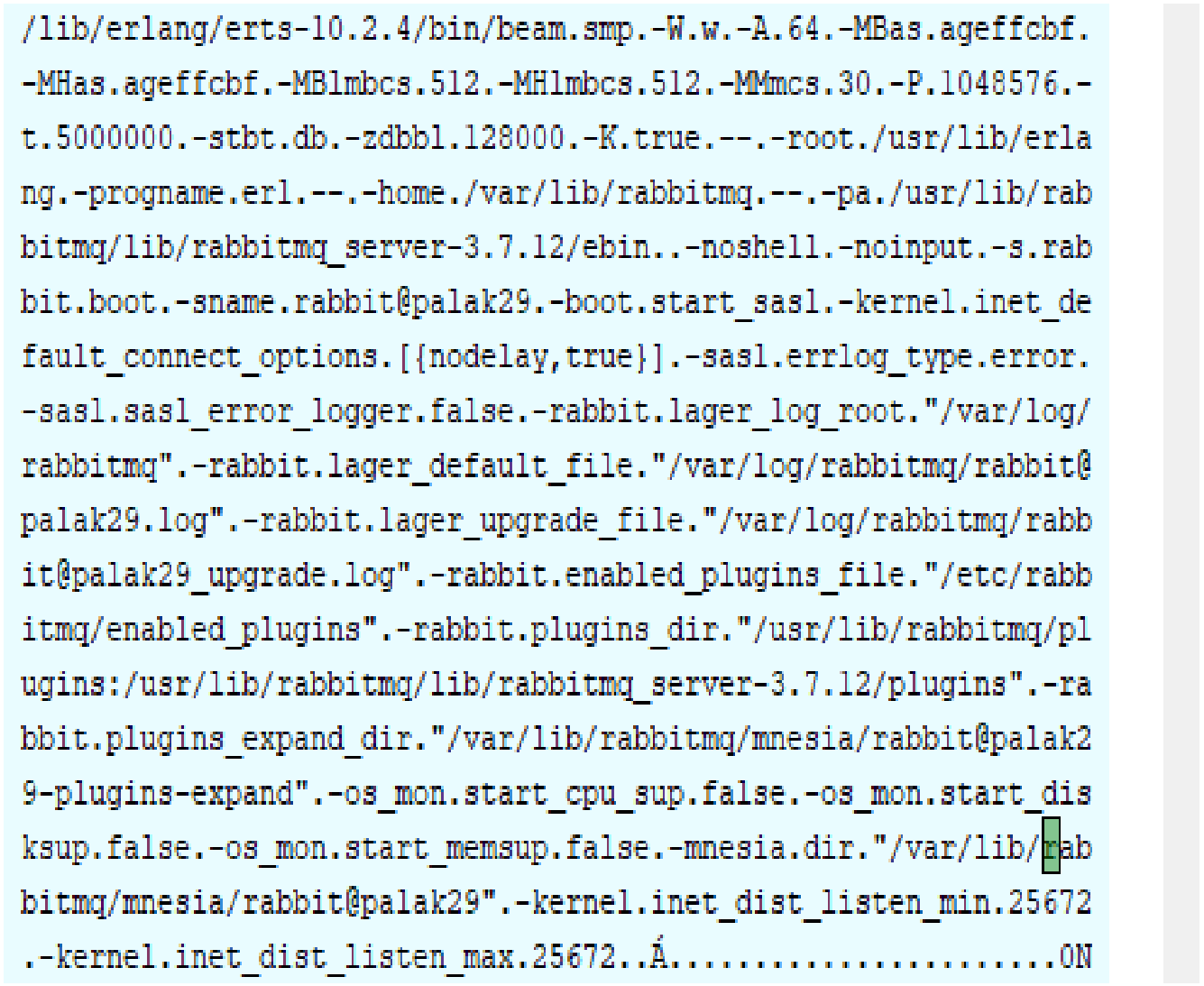

Fig 3. Rabbit MQ broker memory dump example

3. Information such as ip address and path of execution of Mosca broker as well as hostname can also be obtained.

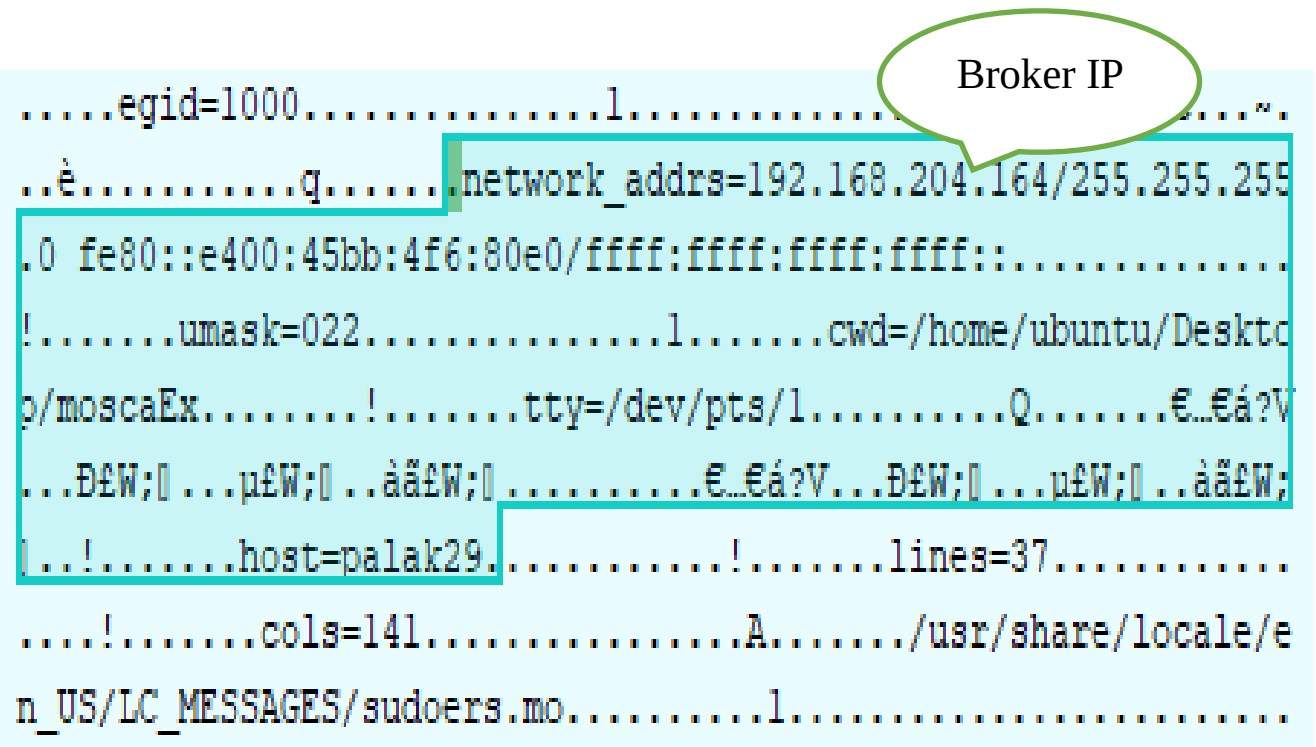

Fig 4. Mosca broker path execution and IP extraction example

4. We can also get the message sent by the subscriber as is shown below of a mosquitto client memory dump.

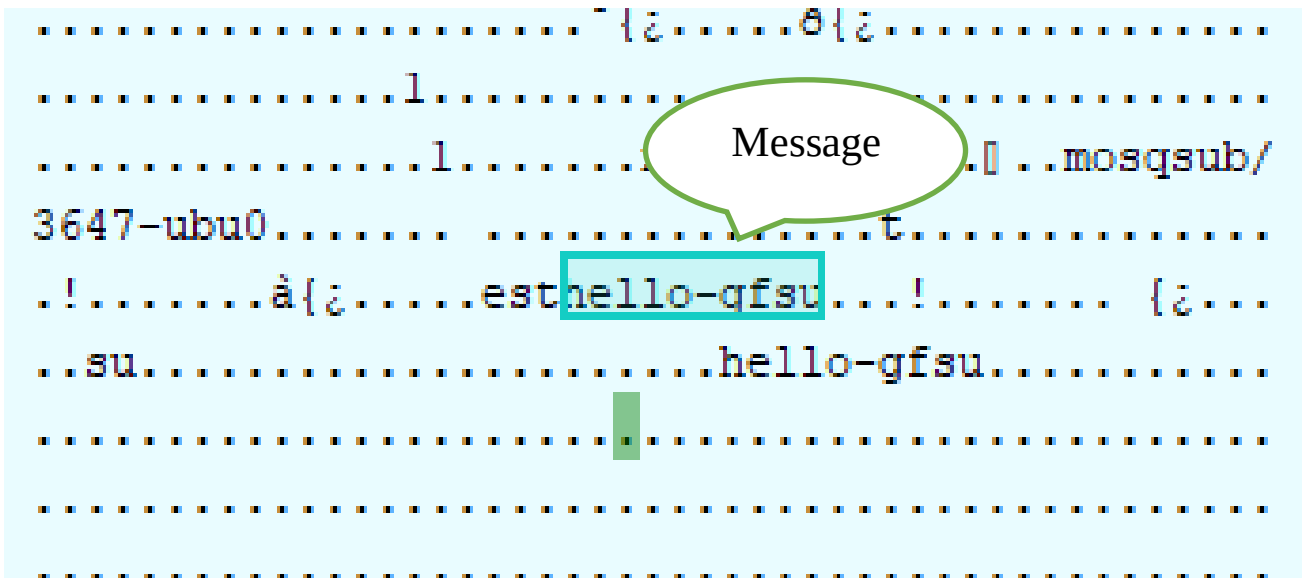

Fig 5. Mosquitto client Message extraction example

5. The time when the Mosquitto broker application started running can also be found.

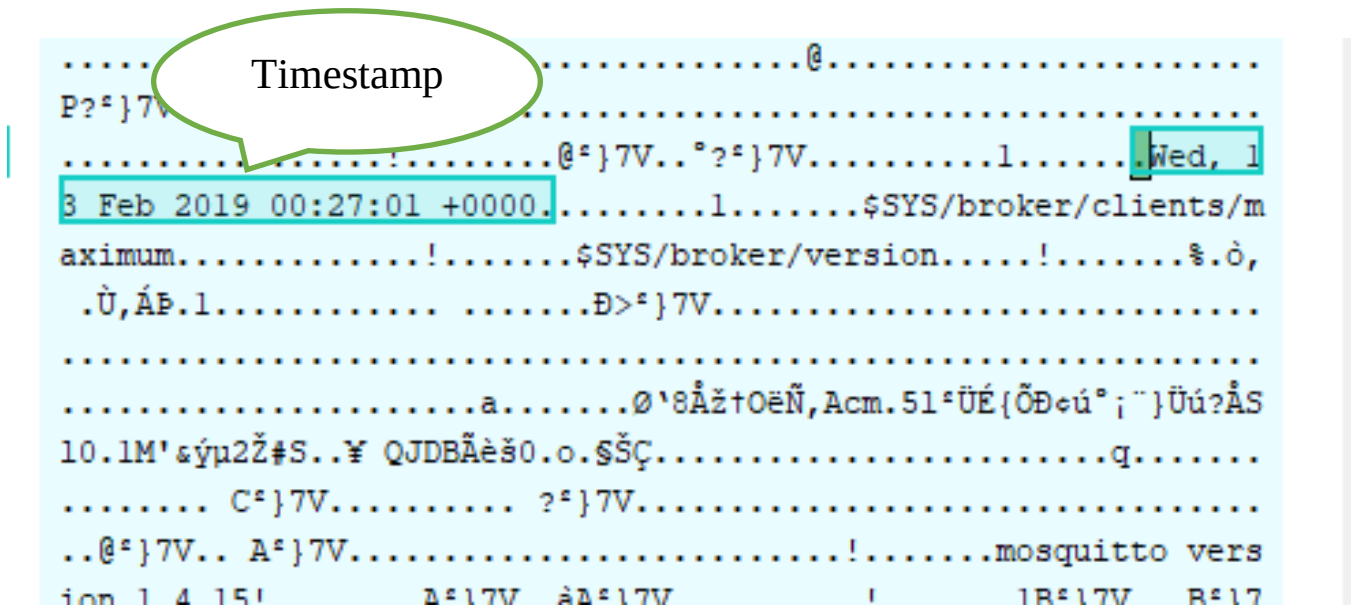

Fig 6. Timestamp captured in mosquito broker

6. The port numbers along with IP addresses of both client and broker can be extracted in the Rabbit MQ broker.

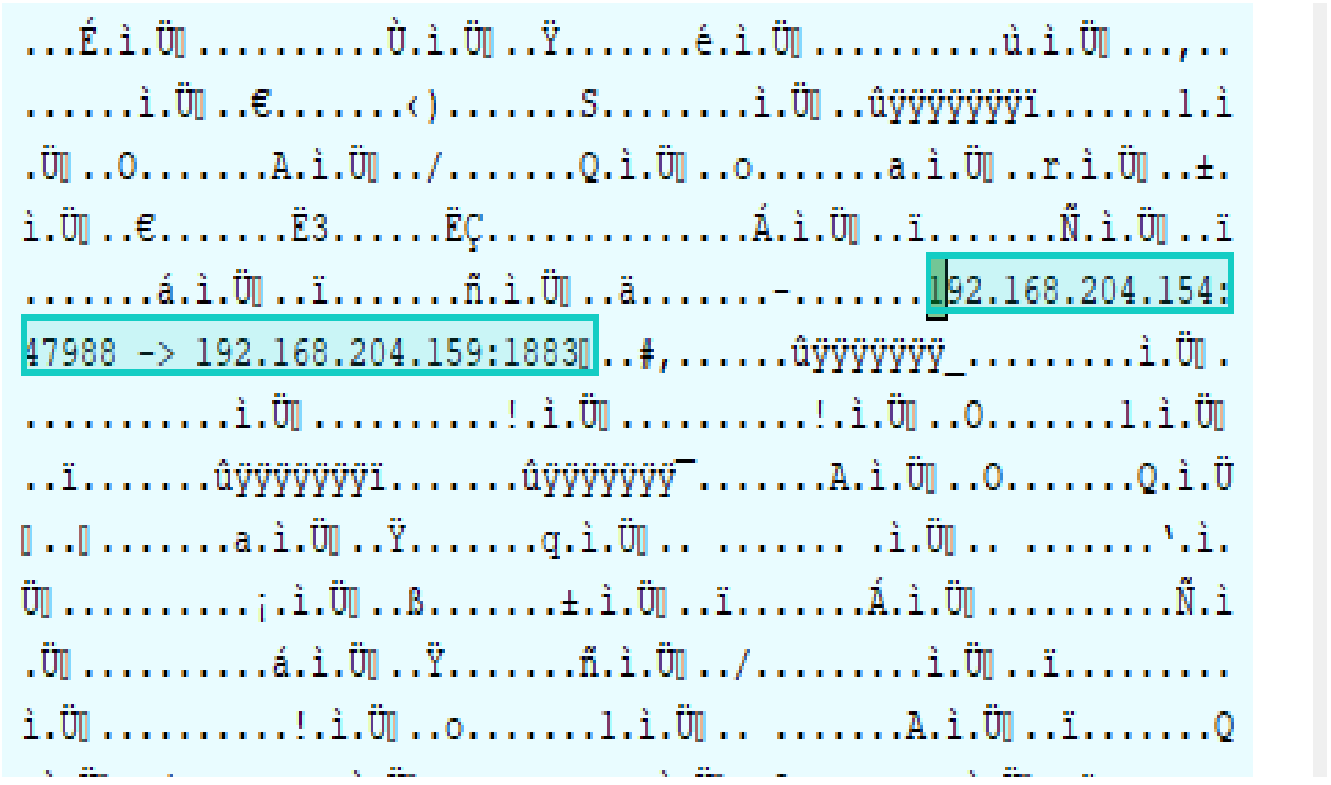

Fig 7. Port and IP addresses captured in Rabbit MQ broker